\documentstyle[aps,epsf]{revtex}

\begin{document}

\title{Spiral Waves in Chaotic Systems}
\author{Andrei Goryachev and Raymond Kapral\\
Chemical Physics Theory Group, Department of
Chemistry, \\ University of Toronto, Toronto, ON M5S 1A1, Canada}

\maketitle
\begin{abstract}
Spiral  waves  are  investigated  in  chemical  systems  whose
underlying  spatially-homogeneous  dynamics  is  governed by a
deterministic   chaotic  attractor.  We  show  how  the  local
periodic  behavior  in  the  vicinity  of  a  spiral defect is
transformed  to  chaotic  dynamics  far  from  the defect. The
transformation  occurs  by  a  type  of period doubling as the
distance  from  the  defect increases. The change in character
of  the  dynamics  is  described  in  terms of the phase space
flow on closed curves surrounding the defect.
\end{abstract}



Spiral   waves   are  commonly  observed  in  oscillatory  and
excitable  media.\cite{exm}  They  are  often  responsible for
the  patterns  one  sees in chemical systems and can give rise
to  spatiotemporal  states  such as defect-mediated turbulence
whose  erratic  dynamics  is characterized by the creation and
destruction  of  pairs  of  defects  (spiral  wave cores) with
opposite   topological   charge.\cite{cgl}   The   topological
charge  $n_t$  is  defined by\cite{mermin} \begin{equation} {1
\over   2   \pi}   \oint  \nabla  \phi({\bf  r})  \cdot  d{\bf
l}=n_t\;,  \end{equation}  where  $\phi({\bf r})$ is the local
phase   and  the  integral  is  taken  along  a  closed  curve
surrounding the defect.

In  this  article  we  examine  the nature of such spiral wave
states  in  chemical  media  where  the underlying dynamics is
itself  chaotic.  More specifically, we consider systems where
the  dynamics  of  the spatially homogeneous system, described
by   ordinary  differential  equations,  has  a  deterministic
chaotic  attractor  which  arises  through  a  period-doubling
cascade.  Consequently,  the  simplest models for the dynamics
considered  here  require at least three phase space variables
in  contrast  to the two-variable descriptions of excitable or
oscillatory   media.\cite{exm}   We  examine  how  this  local
deterministic   chaos   can   support   spiral  waves  in  the
spatially-distributed  medium.  We  also  show that systems of
this   kind   can   exhibit   defect-mediated  turbulence  and
demonstrate  that  the  underlying  local temporal dynamics is
quite different from that in simple oscillatory media.

Consider the Willamowski-R\"{o}ssler \cite{wr} reaction-diffusion equations,
\begin{eqnarray}
{ \partial c_x({\bf r},t) \over \partial t} &=&\kappa_1   c_x   -\kappa_{-1}   
c_x^2   -\kappa_2   c_x   c_y  
+\kappa_{-2} c_y^2 -\kappa_4 c_x c_z +\kappa_{-4} 
+D \nabla^2 c_x\;, \nonumber \\
{ \partial c_y({\bf r},t) \over \partial t} &=&\kappa_2c_x   c_y   
-\kappa_{-2}  c_y^2   -\kappa_{3} c_y + \kappa_{-3} 
+D \nabla^2 c_y\;,\nonumber \\
{ \partial c_z({\bf r},t) \over \partial t} &=&-\kappa_4 c_x c_z +
\kappa_{-4} +\kappa_5 c_z -\kappa_{-5} c_z^2 
+D \nabla^2 c_z\;,
\label{r-d}
\end{eqnarray}
where  $c_{\tau}({\bf  r},t)$  is  the  local concentration of
species  $\tau=x,y,z$  (we  have  suppressed  the arguments of
$c_{\tau}$   on  the  right  hand  side  of  (\ref{r-d}))  and
$\kappa_{\pm  i}$  are  rate  coefficients  that  contain  the
concentrations  of  species  that  are  fixed  to maintain the
system  out  of equilibrium. The diffusion coefficients of all
three species are equal to $D$.

Suppose  the  system is spatially homogeneous and the dynamics
is  described  by the ordinary differential equations based on
the  reactive  terms in (\ref{r-d}). The resulting mass action
rate  law  supports  a  chaotic  attractor  that  arises  by a
period-doubling  cascade.\cite{wr,wu}  The  chaotic  attractor
in  the  $(c_x  c_y  c_z)$ phase space is oriented so that its
projection  onto  the  $(c_x,c_y)$  plane clearly exhibits the
(folded)  phase  space  flow  around  the unstable fixed point
(focus)  ${\bf  c}^*=(c_x^*,c_y^*,c_z^*)$  which  spawned  the
attractor.  Consequently,  to define the phase angle $\phi$ we
change   variables   from   ${\bf   c}=(c_x,c_y,c_z)$   to   a
cylindrical  coordinate  system  $(\rho, \phi, z)$ with origin
at  ${\bf  c}^*$  and  $z$ directed along $c_z$.\cite{sphp} As
the   system   undergoes   a   sequence   of   period-doubling
bifurcations  $\phi$  increases by $2\pi 2^n$ with each period
of  the  oscillation,  where  $2^n$, $n = 1,2,3 \ldots\infty$,
is   the   periodicity   of   the   attractor.   For  period-1
oscillations   the   other   two  variables  can  be  uniquely
parametrized   by   the  phase  $\phi$,  $\rho=\rho(\phi),  \,
z=z(\phi)$  but  this is no longer true after the first period
doubling  bifurcation.  However,  this  variable  suffices for
the   determination   of   the   location   and  charge  of  a
topological defect in the spatially-distributed medium.

\begin{figure}[htbp]
\begin{center}
\leavevmode
\epsfysize = 4in
\epsffile{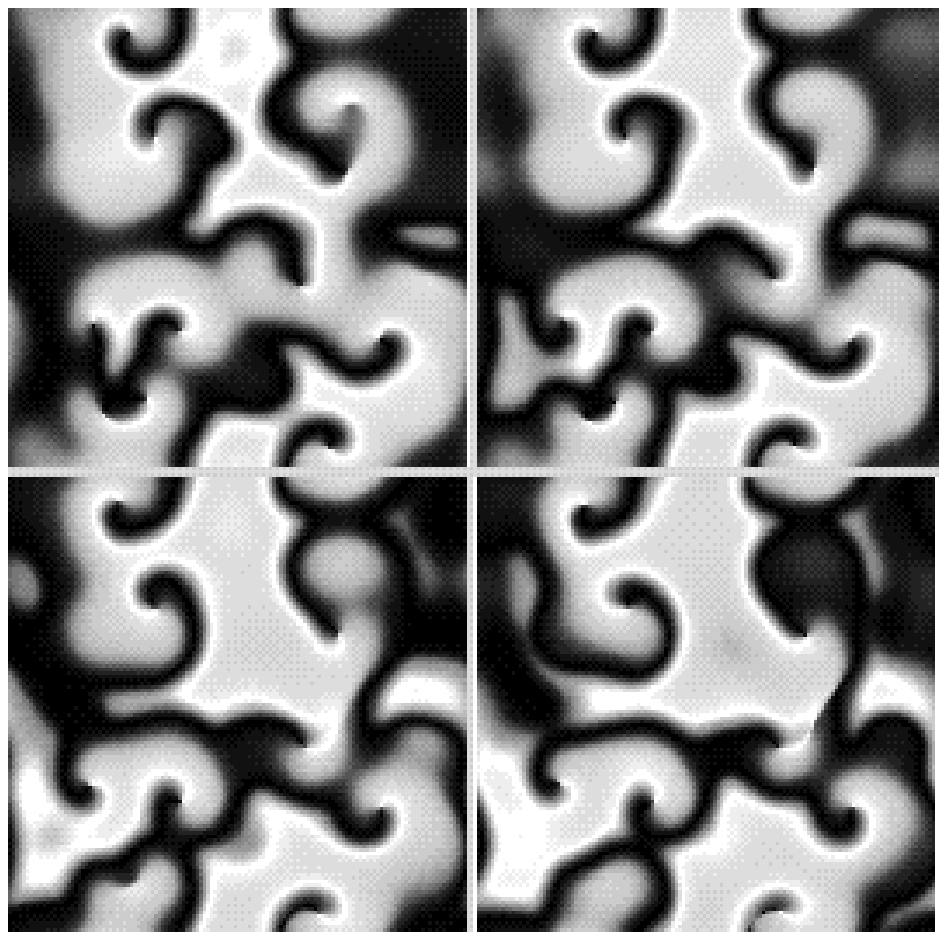}
\end{center}
\caption{Defect-mediated  turbulence  in a chaotic medium. The
local  phase  $\phi(x,y,t)$  is  shown  as  gray shades in the
spatial  $xy$  plane. Defects can be located as the termini of
the   white,   equiphase  contour  lines.  The  time  interval
between  frames  corresponds  to  one  period  of  the  spiral
rotation.  Time  increases  from  left  to  right  and  top to
bottom.      Rate     parameters     are:     $\kappa_1=31.2$,
$\kappa_{-1}=0.2$,    $\kappa_{2}=1.572$,   $\kappa_{-2}=0.1$,
$\kappa_{3}=   10.8$,  $\kappa_{-3}=0.12$,  $\kappa_{4}=1.02$,
$\kappa_{-4}=0.01$,    $\kappa_{5}=16.5$,   $\kappa_{-5}=0.5$.
The  integration  time  step is $\Delta t=5\times 10^{-4}$ and
the  scaled  diffusion  coefficient  is  $D  \Delta t/ (\Delta
x)^2 = 10^{-3}$. Periodic boundary conditions are used.}
\label{dmt}
\end{figure}

Next  we  consider  the  spatially-distributed chaotic system.
Figure~\ref{dmt}   is   a   plot  of  the  local  phase  angle
$\phi({\bf  r})$  in  a  two-dimensional  medium  obtained  by
numerically  integrating  (\ref{r-d}).\cite{numint} One sees a
complex  pattern  of  spiral  defects whose number varies with
time.  All  defects  have  topological charge $n_t=\pm 1$. The
system  evolved  from an initial state with a single defect in
the  center  of the system \cite{incondt}. In the early stages
of  the  evolution the overall number of defects grows rapidly
and  then,  depending  on the ratio of the diffusion length to
the  system  size,  it  either saturates monotonously or rises
to  a  maximum  and decreases to some stationary average value
about  which  it  fluctuates. Note the different rates of evolution
in  certain  parts  of  the  medium.  While the dynamics has an 
almost  periodic  character  in  the  regions  subject  to the
organizing   influence   of  large,  well-established  spirals
(upper  left  and  lower  right  corners of the panels) the  
evolution is  much  faster  in   domains  where 
vortex-antivortex  birth and annihilation take place (the
vortex-antivortex pair seen in the lower left corners of the 
first two panels disappears as time increases).  From  visual  
inspection  of  the  individual
snapshots  of  the  local phase in this figure it is difficult
to  detect  differences  between  this type of defect-mediated
turbulence   and  that  in  oscillatory  media.  Nevertheless,
fundamentally  different  kinds  of  local dynamics consisting
of   perturbed   period-doubled   cycles  and  chaotic  motion
underlie  and  influence the dynamics of the spiral structures
seen in this figure.

The  very  fact  that  stable  spiral  waves exist in a medium
with   underlying   chaotic  dynamics  demonstrates  that  the
reaction-diffusion  kinetics  in  the  vicinity  of the spiral
centers  is  by  no  means  chaotic  --  the  spiral  dynamics
locally   suppresses   the   chaos.  \cite{cml}  However,  the
behavior  in  those parts of the medium which are sufficiently
far  from  any  topological  defect shows neither temporal nor
spatial order.

\begin{figure}[htbp] 
\begin{center}
\leavevmode
\epsffile{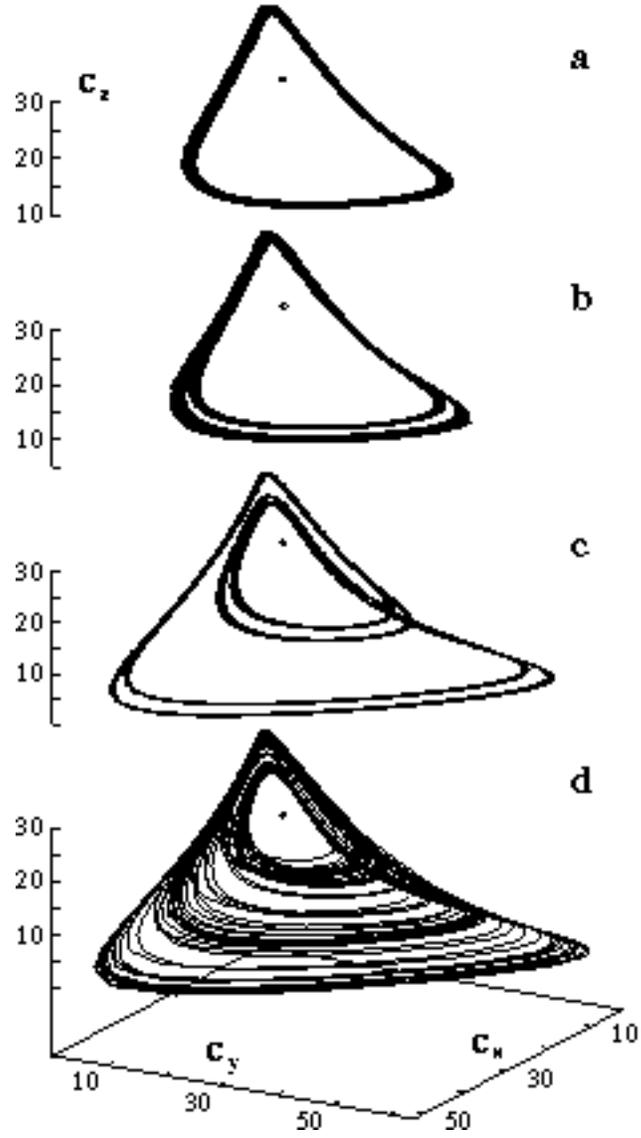}
\end{center}
\caption{Phase  space  trajectories  ${\bf c}(r,\theta,t)$ for
$\kappa_{2}=1.567$   and  the  other  rate  parameters  as  in
Fig.~\protect\ref{dmt}  at  fixed  $\theta=3/4\pi$  for  three
values  of  $r$:  (a)  period-1  limit cycle at $r=0.194$; (b)
period-2   orbit  at  $r=0.206$  and  (c)  period-4  orbit  at
$r=0.344$.  The  scaled  diffusion coefficient is $D \Delta t/
(\Delta  x)^2  =  0.01$  and  no-flux  boundary conditions are
used.   (d)   Local   chaotic   dynamics   at   a   point   in
Fig.~\protect\ref{dmt} far removed from a defect.}
\label{rvar}
\end{figure}

Thus,    we    must    consider   how   the   local, period-1, 
spatio-temporal  dynamics  near  a  defect is transformed into
complex  chaotic  dynamics  far  from  a  defect.  For this
purpose  we  now  examine  the  local  dynamics in a reference
frame  that  is  centered on one of the defects. We denote the
time-dependent  position  of  a  defect  in  the $xy$ plane as
${\bf   r}_d(t)$   and  work  in  a  polar  coordinate  system
$(r,\theta)$   centered   on   ${\bf   r}_d(t)$.   The   local
concentration   fields   may   now   be   expressed  in  these
coordinates:   ${\bf   c}(r,\theta,t)$.  To  investigate  this
local  dynamics  in detail we choose the diffusion coefficient
to  be  sufficiently large and use no-flux boundary conditions
so  that  the  medium  supports  a  single  spiral  wave  that
persists  for  long periods of time. Panels (a)-(c) of Fig.~\ref{rvar} 
show   trajectories  in  the  concentration  phase
space,  ${\bf  c}(r,\theta,t)$,  for  several  values  of  $r$
along  a  ray  emanating  from the defect oriented at an angle
$\theta=3/4\pi$.  One  observes  a period-doubling progression
from  a  perturbed  period-1  limit  cycle near the defect, to
perturbed  period-2  and  period-4  attractors as the distance
increases.   \cite{pert}   This   basic   pattern   of  period
doublings   is  observed  for  all  angles  but  the  circular
symmetry  of  the system is not maintained. The origin of this
effect  will  be  discussed  below when the dynamics on closed
paths surrounding the defect is examined.

Variants  of  this phenomenon were seen in the defect-mediated
turbulent   regime   when  the  dynamics  was  observed  in  a
coordinate  frame  centered  on  a  moving  defect. Far enough
from  the  defect,  locally  one  finds  chaotic dynamics (cf.
Fig.~\ref{rvar}  (d))  which  is suppressed in the vicinity of
the  defect.  Viewed  from the defect, as $r$ increases, chaos
appears  by  a  truncated  period-doubling cascade, similar to
that  seen  in  stochastically perturbed versions of flows and
maps  with  no  spatial  degrees  of freedom.\cite{npd} We are
now  faced  with  the  question  of  how  the locally periodic
dynamics  can  transform  to  locally  chaotic dynamics as $r$
varies   and   maintain  the  spiral  wave  structure  in  the
vicinity of the spiral core.

The  nature  of  this  transformation  can be deduced from the
temporal  behavior  of  ${\bf  c}(r,\theta,t)$  for some fixed
value  of  $r$,  say  $r'$. Suppose the defect has $n_t=+1$ so
that  the  phase  $\phi$  must  change  by  $2\pi$ as $\theta$
varies  through  $2\pi$  and  $r'$  is  such  that  the  local
dynamics  at  $(r',\theta)$ is a period-$2^n$ attractor. Thus,
even  though  the  phase  space  trajectory  of  an individual
point  on  the  circle  with radius $r'$ may be a period-$2^n$
attractor,    the    set    of    points    ${\cal   S}=\{{\bf
c}(r,\theta,t)\;     :\;     0     \le     \theta     \le    2
\pi,\;r=r',\;t=t_1\}$   must   form   a  closed  curve  ${\cal
S}={\cal  S}({\bf  c})$ which loops once around ${\bf c}^*$ in
the  ${\bf  c}$  phase space. The curve ${\cal S}$ cannot span
the  period-$2^n$  attractor  which  loops  ${\bf  c}^*$ $2^n$
times  in  the  course of a cycle. However, ${\cal S}$ deforms
as  time  evolves  in  such  a manner that the trajectory of a
point  on  ${\cal S}$ may be that of a period-$2^n$ orbit, but
the  continuity  of  ${\cal  S}$  is maintained throughout the
evolution.

\begin{figure}[htbp] 
\begin{center}
\leavevmode
\epsfysize = 6in
\epsffile{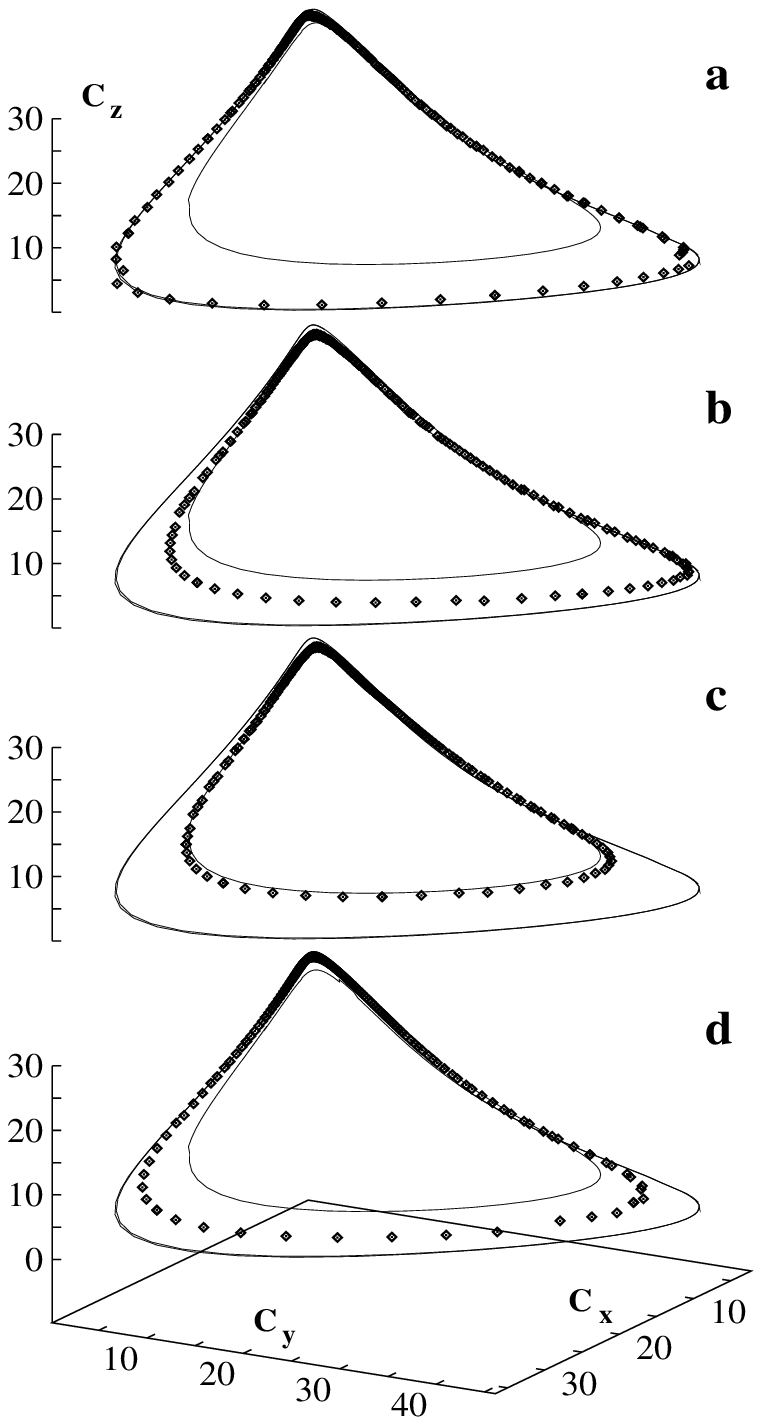}
\end{center}
\caption{Plots  of  ${\cal  S}$  in  the ${\bf c}$ phase space
for  $r=0.219$  for four different times $t_i$: (a) $t_1=0.0$;
(b)  $t_2=0.209$;  (c)  $t_3=0.326$  and (d) $t_4=0.721$; time
is  expressed  in the fractions of average time period of P-2.
${\cal  S}$  is shown as filled diamonds while the solid curve
represents   one   period  of  the  P-2  orbit  at  $(r=0.219,
\theta=3/4\pi)$.  The  simulation  conditions  are the same as
in Fig.~\protect\ref{rvar}.}
\label{thetvar}
\end{figure}

To  make  the  nature of this deformation of ${\cal S}$ clear,
consider   Fig.~\ref{thetvar}  which  shows  ${\cal  S}=\{{\bf
c}(r,\theta,t)\;  :\;0  \le  \theta \le 2 \pi,\;r=0.219,\;t\}$
for  $t=t_1,  t_2,t_3,  t_4$  where  $r$  is  measured  in the
fractions  of  the  system  size. For $r=0.219$ and almost all
$\theta$  the  local attractor is a period-2 limit cycle (P-2)
with   a  small  inner  loop  and  a  large  outer  loop.  For
reference,  one  period  of  P-2  is  shown  in  each panel of
Fig.~\ref{thetvar}   as   a   light  solid  line.  The  filled
diamonds  are  points  on  ${\cal  S}$.  One  sees that at any
fixed  time  instant,  $t=t_i$,  ${\cal S}$ is a simple closed
curve  in  the  ${\bf  c}$  phase space. This curve deforms as
time  progresses  as  follows:  In \ref{thetvar}(a) ${\cal S}$
is  a  large  closed curve that lies on the outer loop of P-2.
As  time  increases  (b)  ${\cal S}$ deforms so that its upper
left  portion  lies  on  the  inner  loop  of P-2 and smoothly
joins  to  that  portion  of ${\cal S}$ remaining on the outer
loop.  In  panel  (c)  at time $t_3$ one sees that deformation
is  complete  and  ${\cal  S}$  lies  entirely along the inner
loop  of  P-2.  Finally,  in  (d),  ${\cal S}$ expands so that
when  the  expansion is complete it again lies along the outer
loop  of  P-2  as in panel (a). Analogous but more complicated
versions  of  such  deformations  are seen for higher periodic
orbits.

We  have  observed  above  that  in the course of the dynamics
the  curve  ${\cal  S}$  joins portions of the inner and outer
loops     of    the    P-2    attractor    (panel    (c)    in
Fig.~\ref{thetvar}).   In   view  of  the  continuity  of  the
medium,  this  implies  that there must be a point on a circle
with  $r=r'$  where  the period-2 bands merge and the dynamics
becomes   effectively   period-1.   This   accounts   for  the
above-mentioned  broken  circular  symmetry  and the fact that
the   period-doubling  progression  observed  along  rays  may
exhibit  different  characteristics  due  to  band  merging at
certain angles $\theta$.

As  one  moves  from  the  vicinity  of  a  defect to the bulk
medium  the  nature  of  the  spatiotemporal  dynamics  may be
summarized   as   follows:  Close  to  the  defect  the  local
dynamics  is  approximately  periodic  and  as $r$ increases a
critical  value  of  $r$  is  reached where the local dynamics
undergoes   a   ``bifurcation"   to   period-2.  This  process
continues  until  a chaotic attractor is obtained. As in noisy
maps   and   flows   this   spatial  period-doubling  sequence
truncates  at  some  finite value beyond which a noisy chaotic
attractor  is  found.  The  number  of  noisy period doublings
that  may  be observed is a function of the system parameters,
for    example    the    diffusion    coefficient,   and   the
characteristic  distance  between  defects or between a defect
and the boundaries.

The  phenomena  described  here  should  exist  in  any system
exhibiting   a   period-doubling  sequence  to  chaos  and  be
experimentally    observable,   for   example,   in   chemical
reactions    carried   out   in   continuously-fed   unstirred
reactors.\cite{cfur}  If  conditions  are adjusted so that the
spatially-homogeneous  system  supports  a  chaotic attractor,
well       known       in       the       Belousov-Zhabotinsky
reaction\cite{bzchaos},   then  suitable  initial  conditions,
similar  to  those  that  are commonly used to initiate spiral
waves  in  excitable  media,  should  produce  the spiral wave
states   described   here.   Typically,   in   experiments  on
well-stirred  systems,  the  period doubling sequence is often
difficult  to  resolve  since  it  occurs  in  a  very  narrow
parameter   range.   This  does  not  imply  that  the  period
doubling  that  occurs  as  one  moves away from the defect is
confined   to   a   narrow  spatial  domain.  To  observe  the
phenomena  described  in  this  paper  it is only necessary to
place  the  system  in  the  chaotic  regime,  or near it. The
spiral  structure  will  locally organize the dynamics and the
passage  to  the  chaotic  dynamics far from the defect should
be observable.

The  work  suggests  the possibility of a variety of phenomena
whose   existence  depends  on  at  least  three  phase  space
variables.  Questions  concering the nature of defect dynamics
and interactions in chaotic media remain to be explored.

This  work  was  supported in part by a grant from the Natural
Sciences  and  Engineering  Research Council of Canada an by a
Killam Research Fellowship (R.K.).

\newpage

\newpage

\newpage


\begin{thebibliography}{99}

\bibitem{exm} A.S. Mikhailov, {\em Foundations of Synergetics I.
Distributed Active Systems}, (Springer-Verlag, Berlin, 1994); 
{\em Chemical Waves and Patterns}, eds. R. Kapral and K. Showalter,
(Kluwer, Dordrecht, 1995).

\bibitem{cgl} P. Coullet, L. Gil and J. Lega, Phys. Rev. Lett. {\bf 62}, 
161 (1989); Physica D {\bf 37}, 91 (1989).

\bibitem{mermin} N.D. Mermin, Rev. Mod. Phys. {\bf 51}, 591 (1979).

\bibitem{wr} K.-D. Willamowski and O.E. R\"{o}ssler, Z. Naturforsch. {\bf
35a}, 317 (1980).

\bibitem{wu} X.-G. Wu and R. Kapral, J. Chem. Phys. {\bf 100}, 5936
(1994).

\bibitem{sphp} One may also use a spherical polar coordinate system
centered on ${\bf c}^*$ in which case the azimuthal angle can be used to
define the phase.

\bibitem{numint} The simulations were carried out using both explicit
Euler methods with time and space steps of $\Delta t=10^{-3}- 10^{-4}$ 
and $\Delta x=10^{-2}- 2\times 10^{-3}$, respectively, as well as split-step 
Fourier transform methods for periodic and no-flux boundary conditions. 

\bibitem{incondt} Initial conditions were chosen to favor the formation of a 
topological defect. The $c_x({\bf r})$ and $c_y({\bf r})$ 
concentrations were varied to produce spatially orthogonal gradients while 
the $c_z$ concentration was fixed at $c_z({\bf r})=c_z^*$, so that
a defect was introduced in the center of the spatial domain. It is 
possible to choose smoothly varying phase conditions so no defects form;  
thus, the defect-mediated turbulent state coexists with simple phase
turbulence.

\bibitem{cml} Such behavior is seen in coupled map lattices with chaotic 
elements [L. Brunnet, H. Chat\'{e} and P. Manneville, Physica D, {\bf 78}, 
141 (1994).]

\bibitem{pert} In the spatially-distributed medium the local dynamics is 
always perturbed by its neighborhood. We use the term ``perturbed" 
period-$2^n$ or chaotic orbit to a denote local attractor whose phase space 
probability density is highly peaked along the corresponding 
period-$2^n$ or chaotic attractor. Henceforth, all periodic or chaotic 
attractors are to be understood in this sense, even if the adjective 
``perturbed" is omitted.

\bibitem{npd} J.P. Crutchfield, J.D. Farmer and B.A. Huberman, Phys. 
Rep. {\bf 92}, 45 (1982).

\bibitem{cfur} G.S. Skinner and H.L. Swinney, Physica D {\bf 48}, 1
(1991).

\bibitem{bzchaos} F. Argoul, A. Arneodo, P. Richetti, J.C. Roux and 
H.L. Swinney, Acc. Chem. Res. {\bf 20}, 436 (1987).

\end{thebibliography}
\end{document}